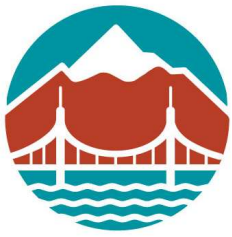

# Parametric modeling of shear wave velocity profiles for the conterminous U.S.

M. D. Sanger[1] and B. W. Maurer[2]

## ABSTRACT

Earthquake ground motions and the related damage can be significantly impacted by near-surface soils. Accurate predictions of seismic hazard require depth-continuous models of soil stiffness, commonly described in terms of shear-wave velocity ($V_S$). For regional-scale studies, efforts to predict $V_S$ remotely, such as the U.S. Geological Survey's National Crustal Model, tend to emphasize deeper lithologic velocity structures, thus simplifying important near-surface soil velocity variations, and tend to be produced at relatively coarse geospatial resolution for one geographic area. In this study, we define a functional form to describe $V_S$-with-depth across the conterminous U.S. We calibrate the parameters of the function using a national compilation of more than 9,000 in-situ geotechnical measurements. By coupling the parametric framework with geospatial machine learning, the model can be leveraged to provide consistent, high-resolution $V_S$-depth predictions of the near-surface geotechnical layer across the U.S., complementing the National Crustal Model and supporting applications such as physics-based ground motion simulations and coseismic hazard assessments.

## Introduction

Earthquake-induced ground shaking is controlled not only by source and path characteristics but also by the near-surface soils through which waves travel. Soft soils can significantly amplify ground motions, prolong shaking, and increase damage to infrastructure, as observed in the 1985 Mexico City and 2011 Tohoku earthquakes. Similar concerns exist in the U.S., for example, in the Pacific Northwest and Atlantic Gulf coastal plains, where deep basins and soft soils may substantially increase earthquake hazards. The most common engineering proxy for site effects is the time-averaged shear-wave velocity ($V_S$) in the uppermost 30 m ($V_{S,30}$). While useful, $V_{S,30}$ simplifies velocity structure into a single number and does not describe depth-dependent gradients that control amplification. Accurate predictions of seismic hazard, particularly for physics-based simulations and site response analyses, require depth-continuous models of $V_S$. For regional-scale studies, or scales at which it is infeasible to obtain continuous measurements, $V_S$ must be predicted remotely.

Towards this end, several regional or "community" velocity models have been developed across the U.S. Examples include models in Utah [1], California [2], the Pacific Northwest [3], the New Madrid Seismic Zone [4], and the Atlantic and Gulf coastal plains [5], among others. Such models are certainly useful, and they often take advantage of high-quality regional geologic information. However, these efforts tend to emphasize deep crustal structure and basin geometry at relatively coarse spatial resolution (~1 km). Near-surface soils, which can vary considerably over tens of meters, are commonly simplified despite their strong influence on site

---

[1] Graduate Student Researcher, Dept. of Civil and Environmental Engineering, University of Washington, Seattle, WA 98195 (email: sangermd@uw.edu)
[2] Professor, Dept. of Civil and Environmental Engineering, University of Washington, Seattle, WA 98195

effects. Similarly at the national scale, the U.S. Geological Survey's National Crustal Model (NCM) [6, 7] represents a decisive step toward a uniform three-dimensional framework (i.e., geospatial predictions of $V_S$-with-depth). However, its current implementation is focused on deep crustal velocity of the western states and provides estimates at 1-km geospatial resolution.

In this study, we establish a parametric model of $V_S$-with-depth that is applicable for the contiguous U.S. We calibrate the functional form using a national compilation of more than 9,000 in-situ geotechnical measurements. The ultimate objective is to establish a parametric framework that can be coupled with geospatial machine learning (ML) for consistent, high-resolution $V_S$-depth predictions of the geotechnical layer across the U.S. By refining the near-surface $V_S$-depth model, this work will complement the NCM, address its near-surface limitations, and support physics-based simulations, site response analyses, and coseismic hazard assessments (e.g., liquefaction and landslides).

## Data and Methodology

The dataset compiled for this study includes $V_S$ profiles, cone penetration test (CPT) profiles, and direct $V_{S,30}$ measurements from publicly available sources across the continental U.S. (Table 1). $V_S$ profiles, measured by downhole, seismic cone, and ReMi methods, are converted to midpoint $V_S$-depth pairs when provided as step-wise velocity layers. CPT profiles are standardized, downsampled to 2.5-m intervals, supplemented with groundwater information where missing [23], and converted to $V_S$ using the average of Andrus et al. 2007 [24] and Robertson 2009 [25] correlations. Direct $V_{S,30}$ measurements are also incorporated to improve spatial coverage. In summation, the collection includes 6,140 $V_S$ profiles and another 3,281 $V_{S,30}$ measurements across the U.S. The locations corresponding to the compiled data are shown in Figure 1.

Table 1. Summary of geotechnical data used in model development.

| Source | | Number of profiles* | Profile type(s) |
|---|---|---|---|
| Ahdi et al. 2017 | [8] | 457 | $V_S$ |
| Sanger and Maurer 2025 | [9] | 445 | $V_S$ |
| Louie 2020 | [10] | 382 | $V_S$ |
| Mohanan et al. 2025 | [11] | 158 | $V_S$ |
| Brandenberg et al. 2020 | [12] | 16 | $V_S$ |
| Mooney and Boyd 2021 | [13] | 11 | $V_S$ |
| Kwak et al. 2021 | [14] | 1793 | $V_S$, CPT |
| Fairbanks et al. 2025 | [15] | 105 | $V_S$, CPT |
| USGS 2019 | [16] | 1057 | CPT |
| Sanger et al. 2024 | [17] | 984 | CPT |
| Rasanen et al. 2024 | [18] | 400 | CPT |
| Rateria et al. 2024 | [19] | 332 | CPT |
| McPhillips et al. 2020 | [20] | 2292 | $V_{S,30}$ |
| Parker et al. 2017 | [21] | 725 | $V_{S,30}$ |
| Utah Geological Survey | [1] | 204 | $V_{S,30}$ |
| Goulet et al. 2014 | [22] | 60 | $V_{S,30}$ |

* Measurements within 0.0001 decimal degrees of one another are dropped to avoid duplicates, with priority given to $V_S$, then CPT, then $V_{S,30}$ measurements, such that the *Number of profiles* is often less than the available measurements in the source database.

With this compiled dataset, the aim is to develop a consistent functional form, or parametric model, to describe the $V_S$-depth relationship in the contiguous U.S. As in other community models, the function is assumed to be monotonically increasing with depth. Several candidate forms are evaluated, including power-law and exponential variants used in prior work (e.g., [2], [5]). Based on general fit suitability identified during preliminary fitting, the adopted parametric form is:

$$V_{S,2.5} + k(z - 2.5)^n = V_S(z) \quad (1)$$

in which $V_{S,2.5}$ is the surface $V_S$ (at 2.5m), $k$ controls the amplitude of the trajectory, and $n$ governs the curvature. Then, a global optimization approach and a differential evolution algorithm are used to explore a wide search space and fit the parameters for each profile. The fitting process optimizes two free parameters, $k$ and $n$, while $V_{S,25}$ is either initialized from the shallowest measured value or, in the case of $V_{S,30}$-only profiles, estimated from an empirical scaling of $V_{S,30}$ derived from preliminary regressions (where $V_{S,2.5}$ is approximately $0.6*V_{S,30}$ in our dataset). During each iteration, the modeled profile is integrated to compute an equivalent $V_{S,30}$, and the loss is minimized to the observed $V_S$ with a penalty term for $V_{S,30}$ misfit. The resulting three-parameter framework of Eq. 1 provides a compact, flexible description of $V_S$-with-depth that is generally transferable across the U.S.

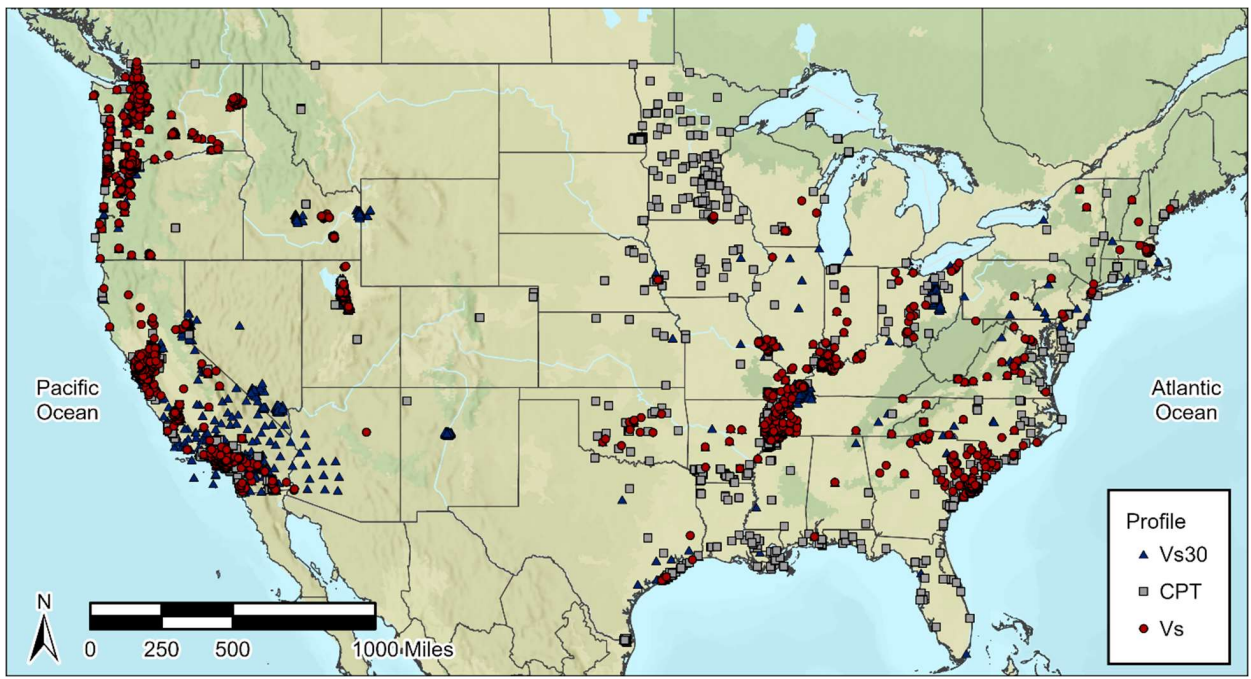

Figure 1. Spatial distribution of geotechnical data used in model development.

## Results and Discussion

The parametric function is calibrated to the compiled dataset, such that each location of in-situ geotechnical measurements has an associated $k$, $n$, and $V_{S,2.5}$ value to describe the near-surface $V_S$-depth relationship. Two example profiles are illustrated in Figure 2, showing (a) a measured $V_S$ profile and (b) an approximate $V_S$ profile derived from CPT data. The profiles are annotated with the fit parameters and $V_{s,30}$, and are shown relative to the NCM queries at that location (i.e., both profiles are from the western U.S.). The differences between the parametric model and the shallow portion of the NCM highlight the opportunity to refine national near-surface velocity models for integration with deeper crustal frameworks.

Recent advances in geospatial modeling provide a clear pathway toward extending this parametric framework to national prediction of $V_S$-with-depth. Zhu et al. 2017 [26] and others demonstrated the power of geospatial information for regional-scale liquefaction hazard predictions, which has since motivated related efforts in coseismic landslide prediction (e.g., [27]) and mechanics-informed ML approaches for liquefaction manifestation (e.g., [28], [29]). In parallel, geospatial ML of many predictor variables has shown to substantially improve $V_{s,30}$, prediction relative to traditional regression approaches based on slope or geology alone [30]. These developments indicate that geospatial proxies, such as topography, surficial geology, and groundwater depth, may effectively capture geospatial patterns in subsurface properties, such as $V_S$-with-depth. Extending the geospatial ML methods to the parametric $V_s$-depth model developed here offers a means to generate high-resolution national products that stand alone as a near-surface $V_s$ model for seismic hazard assessment, and which can complement deeper crustal models such as the NCM. By anchoring predictions in measured data and translating them into a compact, interpretable parameter set (i.e., $k$, $n$, and $V_{S,2.5}$), this model positions itself as a bridge between localized observations and national hazard models, ultimately improving the treatment of near-surface site conditions in U.S. seismic hazard assessments.

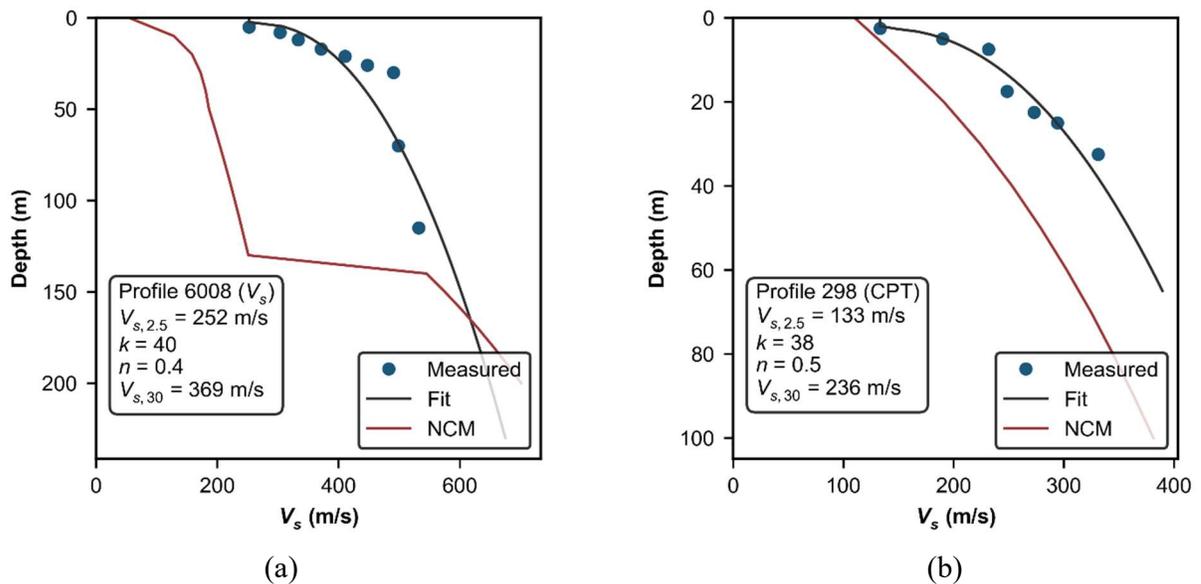

(a)                                              (b)
Figure 2. Example parametric model fits to a measured (a) $V_s$ and (b) CPT profile, with the NCM profile queried at that location.

## Conclusions

This work establishes a nationally consistent parametric model of $V_s$-with-depth for the conterminous U.S. Using more than 9,000 in-situ geotechnical measurements, including $V_s$ profiles, CPTs, and $V_{s,30}$ observations, we calibrate a three-parameter functional form that is designed to capture near-surface gradients and deeper structure while remaining stable and interpretable. The approach is informed by and anchored to measured profiles, when available, as well as the measured $V_{s,30}$. The framework presented here is not intended to replace community velocity models, which often benefit from high-quality region-specific knowledge. Rather, it provides a uniform "background" model of the geotechnical layer at the national scale. By combining the parametric structure with machine-learning prediction of parameters from proxy variables, future efforts can deliver high-resolution $V_s$-depth maps that complement the NCM, improve representation of near-surface soils, and enhance physics-based simulations, site response analyses, and hazard assessments nationwide.

## Acknowledgments

The presented work is based on research supported by the U.S. Geological Survey (USGS) under award G24AP00441. However, any opinions, findings, conclusions, or recommendations expressed herein are those of the authors and may not reflect the views of USGS.